# ECOGEN, an open-source tool dedicated to multiphase compressible multiphysics flows.

Kevin Schmidmayer[(1)], Antoine Marty [(2)], Fabien Petitpas [(2)], Eric Daniel [(2)]

[(1)] *California Institute of Technology, Division of Engineering and Applied Science, Pasadena, CA 91125, USA. kevinsch@caltech.edu*
[(2)] *Aix Marseille Univ, CNRS, IUSTI, Marseille, France.*

## 1. ABSTRACT

This paper presents a new multiphase flow code, cast under an open-source GNU license. The main characteristics of the different flow models are given, then the numerical method used is briefly presented: it includes temporal flow solvers, meshing features (like AMR technics), results visualization.

Two examples of flows solutions are presented: the interaction of a high-speed flow with a droplet and the second concerns the attenuation of a propagating shock wave.

## 2. INTRODUCTION

Many numerical flow models have been developed devoted to rather different flows that concern multiphase flows (1), (2), (3), (4), (5), (6). It appeared necessary to develop a stable numerical platform that includes all those models without losing efficiency in programming and make easier the diffusion in the scientific community. ECOGEN is the tool devoted to this task. It has been developed and distributed under an open-source (GNU GPLv3 license). This code is available online at the following address:

https://github.com/Matshishkapeu/ECOGEN

In this paper we present the basic of the multiphase compressible flow models include in ECOGEN. The main features of the code – considering the temporal solver, the new AMR technics, the domain decomposition – are present. Then, the results concerning two high speed flow results are given: the first one concerns the atomization of a liquid droplet and the second is about the attenuation of a propagating shock wave. In this last case, we also present comparisons with experimental data obtained in our shock tube lab.

## 3. FLOW MODELS

Several compressible multiphase flow models are included in ECOGEN. There are two general common features for these models which are:

- they are based on a diffuse interface method

- they are hyperbolic (strictly or at least nearly).

The first property allows to treat a large variety of multiphase flows including mixtures in equilibrium state as well as interface flows (meaning that each phase is a pure phase from either side of an interface). A control volume of fluid is made of several phases (components). There is no interface reconstruction neither interface tracking. The interface is seen as a phase quantity, generally the volume fraction.

The compressibility property implies the use of equations of states (EOS). Various EOS are included in ECOGEN: perfect gas, stiffened gas. Next developments concern JWL, Mie-Grüneisen EOS and extension to tabular equations of state. Flows governed by Euler Equations are also included in ECOGEN (see paragraph **5.3**).

The multiphase flows are basically described by the Kapila model (7). In this model, each phase evolves with the same pressure and with the same velocity, that is a mechanical equilibrium. Without any transfer between the phases this model is expressed as:

$$\begin{cases} \frac{\partial \alpha_k}{\partial t} + \vec{u}.\vec{\nabla}\alpha_k = K\vec{\nabla}.\vec{u} \\ \frac{\partial \alpha_k \rho_k}{\partial t} + \vec{\nabla}.(\alpha_k \rho_k \vec{u}) = 0 \\ \frac{\partial \rho \vec{u}}{\partial t} + \vec{\nabla}.(\rho \vec{u}\vec{u} + p\bar{\bar{I}}) = \vec{0} \\ \frac{\partial \rho E}{\partial t} + \vec{\nabla}.[(\rho E + p)\vec{u}] = 0 \end{cases} \quad (1)$$

In this system, the quantities with *k* index are related to k-*th* pure phase, with no index theses terms represent mixture quantities. $\alpha_k$ is the volume fraction, other quantities are written with usual notations. Notice that E is the total energy $E = \varepsilon + \frac{\|\vec{u}\|^2}{2}$.

In the volume fraction evolution equation, *K* represents the difference in the compressibility behaviors of each phase: it is linked to the acoustical impedance of pure phase. This term can be analytically expressed, or it can be computed by the means of relaxations terms in other models (8) (9) (10).

ECOGEN also contains a mechanical and thermal equilibrium model. This last model is well designed for mixture of pure fluids problems, meaning that the approach is closed to Direct Numerical Simulation. The system of partial differential equations looks like the multi-components Euler equations:

$$\frac{\partial \alpha_k \rho_k}{\partial t} + \vec{\nabla}.(\alpha_k \rho_k \vec{u}) = \sum_j \dot{m}_{kj}$$
$$\frac{\partial \rho \vec{u}}{\partial t} + \vec{\nabla}.\left(\rho \vec{u}\vec{u} + p\bar{\bar{I}}\right) = \vec{0} \quad (2)$$
$$\frac{\partial \rho E}{\partial t} + \vec{\nabla}.[(\rho E + p)\vec{u}] = 0$$

Even if the temperatures are the same for all the phases, this model allows the treatment of mass transfer between phase *k* and *j* ($\dot{m}_{kj}$) because the Gibbs free energies depend on the phase considered and are not in equilibrium state.

This large number of model in ECOGEN allows to treat a set of compressible flows. Some exchanges between phases can also be considered:

Mass transfer: the models included in ECOGEN enable the treatment of cavitating flows (1). Combustion of a pure phase as well as detonation waves can be modelled (2). Change phase problems such as vaporization or condensation can also be solved.

Capillary effects: a new model proposed in (6) enables a conservative treatment of such problem. An extension of Model (1) is necessary which implies the treatment of the different physics by a numerical splitting. An example is provided in next section, devoted to the droplet fragmentation in a high-speed flow.

Thermal transfers: in ECOGEN are included heat transfers between the compressible phases as well as between a phase and a wall.

Viscosity effects: those effects may appear when momentum exchange occurs between phases and also as a dissipative term for pure phase flows.

## 4. CODE FEATURES

In this section are presented the main features of ECOGEN. These features involve numerical scheme for solving the flow equations as well as computing considerations.

The scheme is based on Godunov's method in a finite volume framework. Fluxes are solved thanks to exact or approximate Riemann solvers (HLLC). A MUSCL type extension provides high order accuracy in space and time.

The code is developed in C++. The parallelization is based on a geometrical domain decomposition and used MPI functions library to ensure the communications between the various domains.

The meshes used are structured or unstructured, can be done in 1-2-3D. ECOGEN can read *gmesh* mesh format for unstructured geometries but can also generate Cartesian grids.

In order to improve the quality of the solution around discontinuities an AMR technics has been developed. A new algorithm ensures efficient numerical solution for unsteady flows and the final code is much easier to develop than with previous existing methods (11). AMR is only available for structured meshes.

The data visualization can be done with VTK tool, output files are provided under XML standard format.

## 5. RESULTS

### 5.1. Introduction

We present here a sample of results obtained with ECOGEN. The first example is the simulation of the atomization of a droplet in a high-speed flow including capillary effects. The second example is about the attenuation of a propagating shock wave in ducts: numerical results are compared with experimental ones.

### 5.2. Atomization of a droplet in a high-speed flow

A mathematical model for multiphase flows with capillary effects that ensures conservation of mass, energy and momentum is presented in reference (6). This model also verifies the second law of thermodynamics for capillary flows. This new model is a wide extension of Kapila model, in which the supplemental physics is numerically solved thanks to an operator splitting involving three different stages.

The first one solves the equations of the multiphase flow with a pressure disequilibrium. In the second stage, the effects of surface tension are solved and finally, the pressure equilibrium between phases is recovered thanks to a relaxation procedure.

The surface tension is modelled via the Brackbill volume force (12) that required a color function $c(\overrightarrow{OM}, t)$ and this force is expressed via the capillary tensor to ensure a conservative form of the equations. It is noticeable that in the new approach, the color function $c(\overrightarrow{OM}, t)$ is no longer a phase quantity such as the volume or the mass fraction and thus the model is shown hyperbolic. To compute the capillary tensor the normal vector to the surface is required: it is computed via the gradient of $c(\overrightarrow{OM}, t)$. The components of the gradient of $c(\overrightarrow{OM}, t)$ obey a conservative equation which is solved.

A 3D numerical solution is now presented. The test case consists in the interaction of a water droplet and a gaseous high-speed flow (that may be a shock wave).

The liquid obeys a stiffened gas equation of state:

$$p = (\gamma_{liq} - 1)\rho\varepsilon - \gamma_{liq} P_{\infty,liq} \qquad (4)$$

with $\gamma_{liq} = 4.4$ and $P_{\infty,liq} = 6.10^8 Pa$.

The air obeys a perfect gas law: $p = (\gamma_{air} - 1)\rho\varepsilon$ with $\gamma_{air} = 1.4$.

The test case is depicted in Figure 1 in which are given some flows data.

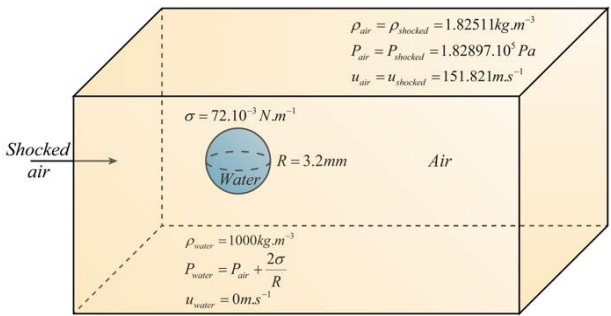

Figure 1: Description of the 3D test case.

An AMR mesh containing initially 250x50x50 cells for a physical domain of (250mm*50mm*50mm) is performed. The AMR method developed in ECOGEN is a new one (11) (13). 4 levels of refinement are used that leads to an equivalent non-AMR mesh size of $2.56*10^9$ cells.

In Figure 2 is shown a view of the mixture density as well as the mesh. One can easily observe the mesh refinement around the droplet and at its rear (around the vortices). This view is plotted at time t=1ms. The initial flow velocity is around 152m/s: this velocity jump is due to an incoming shock wave with a pressure jump equal to 1.83. The high values of the density numerical schlieren obviously correspond to the liquid and appear in red in this plot. One can observe that the droplet is flattened with filaments at its side. These filaments will be torn away and form small droplet during the atomization process. The method can separate filaments from the 'mother droplet' as well as to form very small droplets.

These smaller droplets could be counted in Figure 3, that shows the liquid volume fraction at the same time.

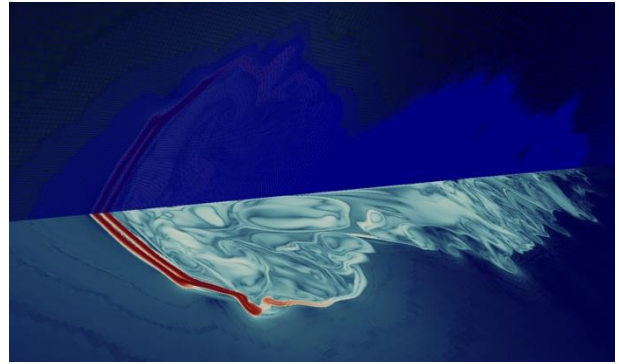

Figure 2: Mesh and mixture density numerical schlieren at time t=1ms.

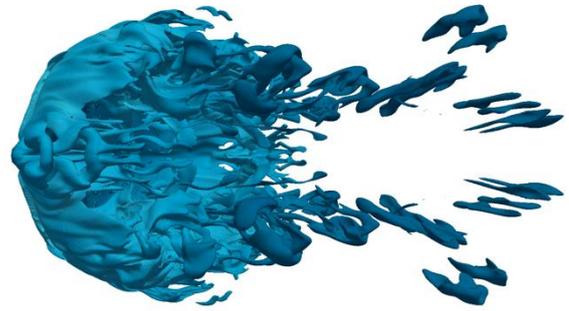

Figure 3: Liquid volume fraction at time t=1ms.

In Figure 4, are compared the CPU execution time if one considers 3 maximal levels of refinement or 4 levels versus the physical time. There is an order of magnitude but with only 3 levels of refinement, the simulation does not provide enough accurate results for this 3D test case, accuracy is considered here as the numerical diffusion at the liquid-gas interface.

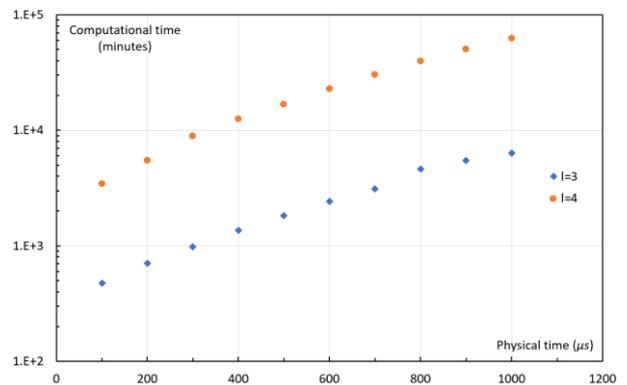

Figure 4: Computational time function of physical time for the 3D atomization tests.

## 5.3. Attenuation of a propagating shock wave

ECOGEN was used in support of an experimental study about the propagation and the attenuation of a shock wave through a 'Y' bifurcation (14) with a trap located in one of the branch (see Fig. 5.) to see what could be the impact on the attenuation of the end-wall reflected pressure. The incident shock wave Mach number was experimentally deduced from pressure records (PCB 113B26) taken at two different locations along the shock tube wall. In the present case the incident Mach number is 1.44 and moves from the left to the right.

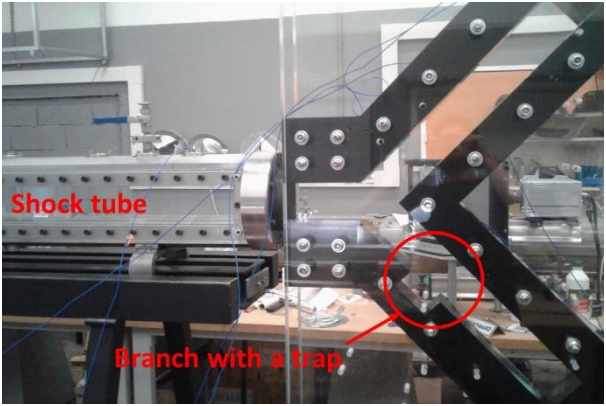

Figure 5: Picture of the shock tube with the 'Y' bifurcation with a trap device.

The physical model solved with ECOGEN is based on a Euler equations solver. Numerical results are now compared to the experimental results both qualitatively (wave patterns in Fig. 6) and quantitatively (pressure signals in Fig.7). Fig.6 presents a comparison of schlieren pictures (experimental on the left versus numerical on the right) taken at same time. The schlieren variable calculated in the present work is the magnitude of the density gradient computed at each cell and visualized using open-source Paraview software. Specifically, the numerical schlieren is calculated here as follows.

$$S = log_{10}(1 + \|\vec{\nabla}\rho\|) \quad (5)$$

In view of the complexity of the present flow in the vicinity of the trap, we can reasonably consider the numerical calculation is in good agreement with the experimental observations. In Fig. 7 comparisons of pressure signals obtained experimentally and numerically for an incident shock wave Mach number of 1.44 are shown for a sensor located on the end-wall of the trapped branch. Fig. 7 shows that no significant difference (less than 5% on the recorded mean value) exists between numerical and experimental pressure traces. The small difference between the two signals is certainly due to the slightly error between the experimental and numerical sensor location. Note that the experimental value of the end-wall reflected pressure is slightly below the numerical one and can be explained by the non-ideal experimental conditions as calibration of sensors, parasitic losses in the device, small leaks or unperfect initial experimental conditions.

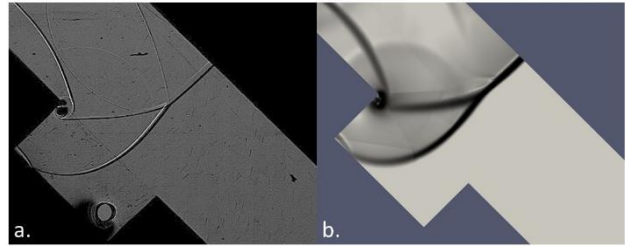

Figure. 6: Experimental (left) and numerical (right) schlieren pictures showing the expansion of a planar shock wave through trapped duct for a Mach number equal to 1.44.

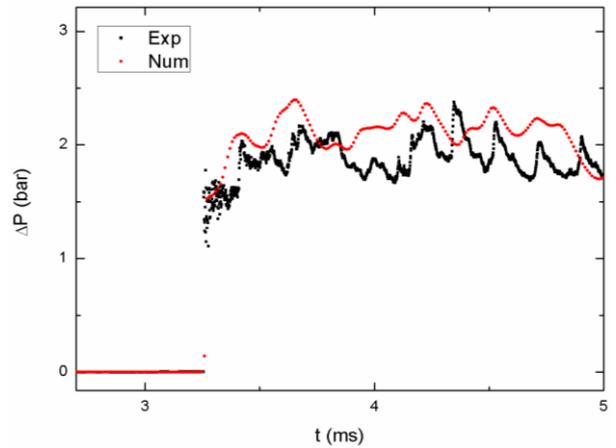

Figure. 7: Comparison of the evolution of the end-wall reflected pressure signal in the Y-shape for an incident shock wave Mach number of 1.44. The experimental (black line) and its numerical equivalent (red line).

The purpose of the numerical support in this experimental study is to complete the experimental investigation of the flow field in the device by adding some main physical mechanisms that only a numerical analysis could afford. As the velocity flow field behind the incident shock wave or the complete mapping of the pressure all along the device and not only at the sensor location. This analysis includes a numerical shadowgraph of the flow field behind the shock which highly depends on the incident Mach number. See Figure 8 for example.

Finally, a numerical parametric study on the size and the location of the trap in order to maximize the effects on the attenuation of this one on the end-wall reflected pressure. The numerical contribution in this case of investigation saves us time and a better design of the experimental setup.

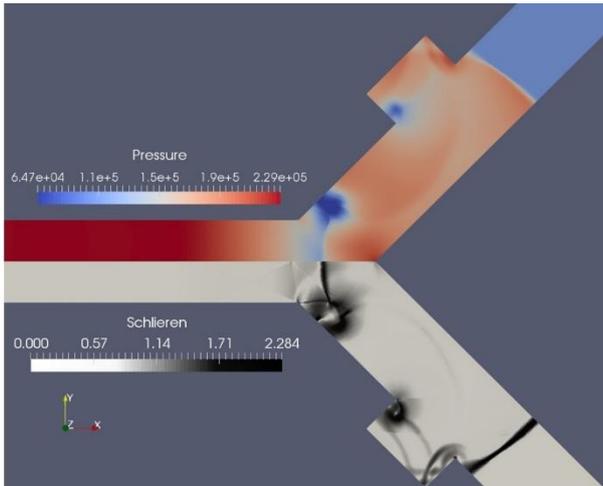

Figure 8: Numerical superposition of the schlieren density (bottom) and the pressure field (top) for Mach number of 1.44.

The time of calculation is about three hours for an area of 0.036m² made up of about 750,000 mesh cells with a labtop HP Zbook G3, 3 processors.

## Références